\documentclass[a4paper,10pt,twocolumn,preprint,3p]{elsarticle}
\usepackage{amsmath,amssymb,graphicx}
\usepackage{caption}
\usepackage{subcaption}
\usepackage{soul}
\usepackage{textgreek}

\makeatletter
\def\ps@pprintTitle{%
  \let\@oddhead\@empty
  \let\@evenhead\@empty
  \let\@oddfoot\@empty
  \let\@evenfoot\@oddfoot
}
\makeatother

\title{Tunable optical parametric generator based on the pump spatial walk-off}

\author[1,2,*]{Andrea Cavanna}
\author[1]{Felix Just}
\author[3]{Polina R. Sharapova}
\author[1]{Michael Taheri}
\author[1]{Gerd Leuchs}
\author[1,2,3]{Maria V. Chekhova}

\address[1]{Max Planck Institute for the Science of Light, G\"unther-Scharowsky-Str. 1/Bau 24, 91058 Erlangen, Germany}
\address[2]{University of Erlangen-N\"urnberg, Staudtstr. 7/B2, 91058 Erlangen, Germany}
\address[3]{Department of Physics, M.V. Lomonosov Moscow State University, Leninskie Gory, 119991 Moscow, Russia}

\address[*]{Corresponding author: andrea.cavanna@mpl.mpg.de}

\begin{document}
\begin{abstract}
We suggest a novel optical parametric generator (OPG) in which one of the down-converted beams is spontaneously generated along the Poynting vector of the pump beam. In this configuration, the generation takes advantage of the walk-off of the extraordinary pump rather than being degraded by it. As a result the generated beams, signal and idler, are bright, due to a high conversion efficiency, spatially nearly single-mode, due to the preferred direction of the Poynting vector, tuneable over a wide range of wavelengths, and broadband. The two beams are also correlated in frequency and in the photon number per pulse. Furthermore due to their thermal statistics these beams can be used as a pump to efficiently generate other nonlinear processes.
\end{abstract}


\maketitle

\section{Introduction}
During the last decades, laser technology aims to cover all possible wavelength ranges, from ultraviolet to infrared and terahertz. Although available laser active materials and, accordingly, available lasing wavelengths are limited, one can still fill almost all spectral gaps with the help of nonlinear optics.
The common tools to generate radiation at a desired frequency are optical parametric oscillators (OPOs) in which the fields oscillate in a cavity, in order to increase the conversion efficiency. This, however, removes the possibility of generating picosecond and shorter pulses. One can use in this case pulsed pump and continuous-wave seeding, but this is technically more complicated.
 Here, we analyse a new OPG, which is based on high-gain parametric down-conversion (PDC) along the pump Poynting vector and therefore requires neither cavity nor seeding. It is wavelength tuneable, spatially single-mode and broadband, with the spectral width easily adjustable.

Phase matching for PDC is often achieved through birefringence, with the pump extraordinary polarised and  therefore subject to spatial walk-off. Inside the crystal, the Poynting vector of the pump is tilted by an angle $\rho$ without affecting the direction of the wavevector. In order to achieve a high conversion efficiency it is preferable to have collinear phase matching, so that the radiation is self-amplified, in combination with tight focussing. Under these conditions, unless two-crystal compensation schemes are used~\cite{Bosenberg:89}, the pump beam walks off from the signal and idler beams and does not amplify them any more. Furthermore, the signal and idler beams can be up-converted to the pump frequency with an overall effect of reducing the spatial beam quality. A common method to prevent the up-conversion is to use non-collinear phase matching~\cite{Liang:07, Tiihonen:04}, but this reduces drastically the interaction length in the crystal. In OPOs, the cavity compensates for the reduced length of interaction.

In our system we generate PDC along the pump Poynting vector, using therefore a non-collinear phase matching, as it is often done in seeded OPOs~\cite{Gale:95}. Since in this configuration the walk-off is not anymore a limitation factor, a long crystal can be used with tighter pump focussing, the only limitation being the Rayleigh length of the pump. Moreover, up-conversion is avoided and neither a cavity nor seeding is needed as the parametric gain is high enough. As we will show, the waist of the pump determines the bandwidth of the amplified signal beam and therefore the corresponding idler one. By tilting the nonlinear crystal one can also modify the phase matching and tune the central wavelength.

\section{Anisotropy effect at high parametric gain}
\begin{figure}
\begin{subfigure}[h]{0.49\columnwidth}
\includegraphics[width=\columnwidth]{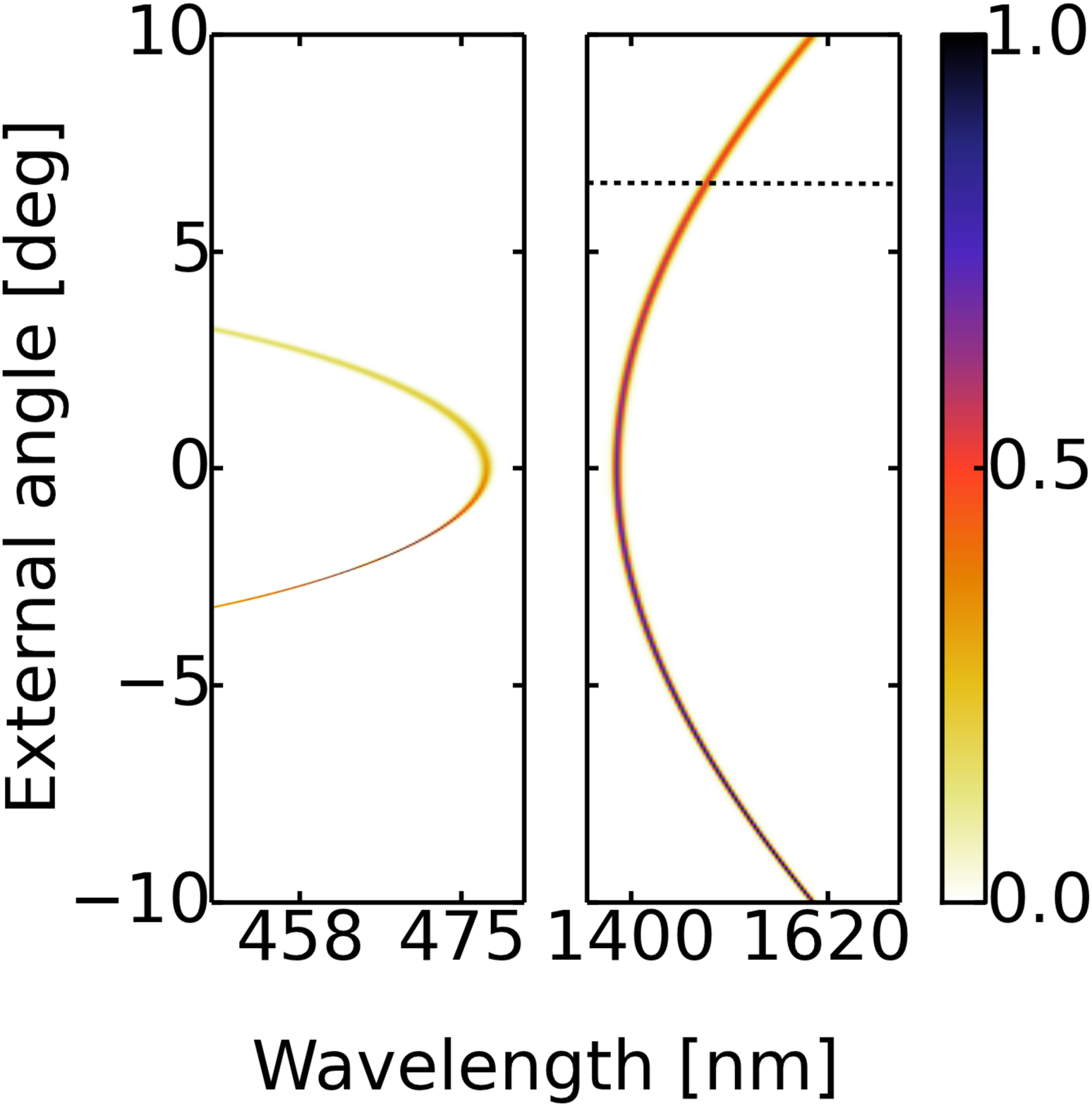}
\caption{$\Gamma = 0.001$.}
\label{fig:SpectralAmplitudeLowGain}
\end{subfigure}
\begin{subfigure}[h]{0.49\columnwidth}
\includegraphics[width=\columnwidth]{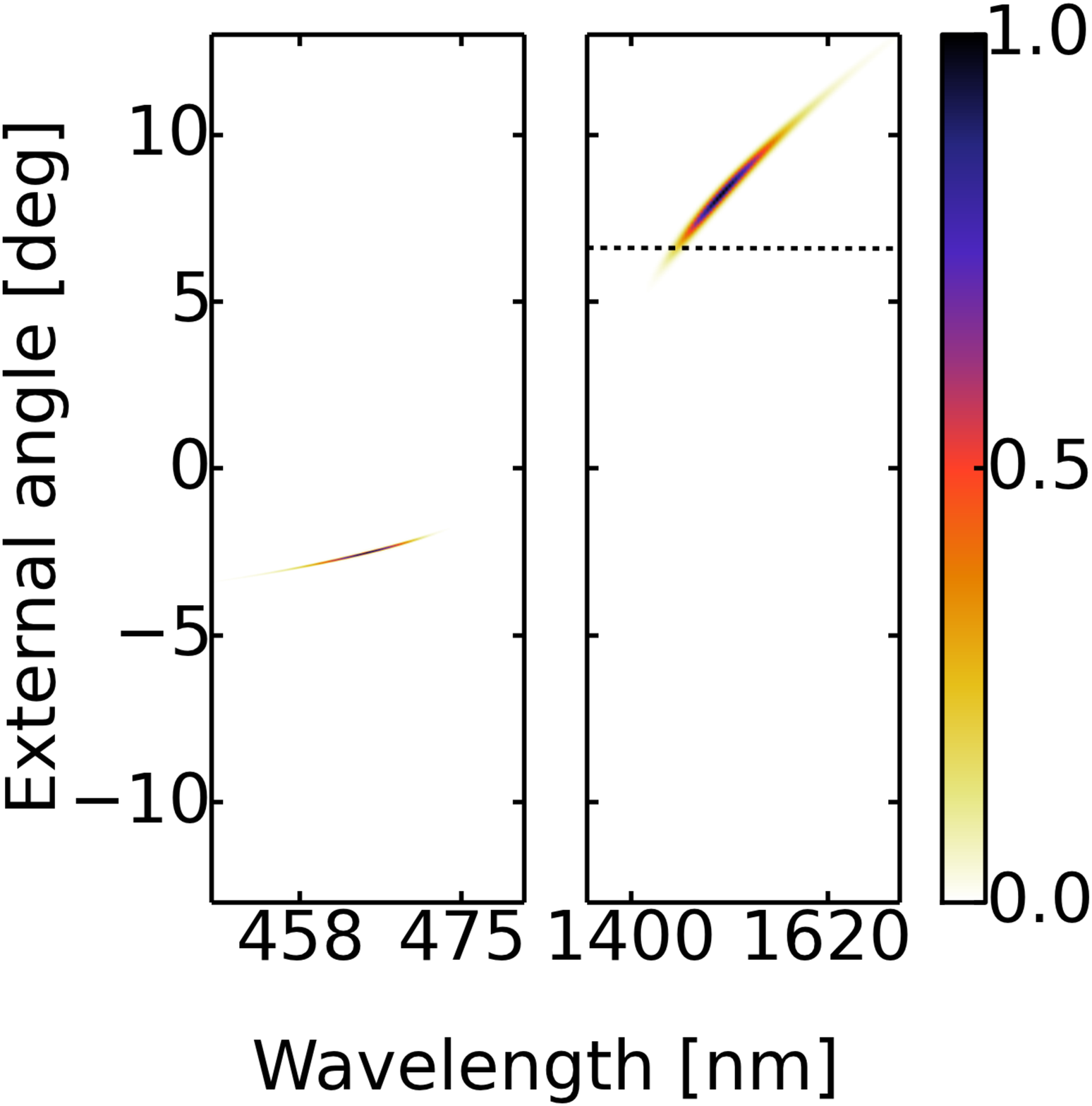}
\caption{ $\Gamma = 50.0$}
\label{fig:SpectralAmplitudeHighGain}
\end{subfigure}
\caption{Wavelength-angular intensity spectrum for low-gain (left) and high-gain (right) PDC.}
\label{fig:SpectralAmplitudeGain}
\end{figure}

There are several approaches to the description of walk-off effects in low-gain PDC, see for example \cite{Fedorov:07, Perina:15}, but at high gain, a different model is needed~\cite{Sharapova:15}. This model considers the transverse wavevector spectrum of PDC at a fixed wavelength, and it is applied to all wavelengths within our range of interest. The key characteristic is the two-photon amplitude (TPA), the probability amplitude for the signal photon to be emitted at angle $\theta_s$ and the idler photon, at angle $\theta_i$. The standard description \cite{Hong:85, Just:13} does not include the effect of the walk-off and therefore cannot be used. The TPA with the pump walk-off taken into account has the form~\cite{Cavanna:14}
\begin{equation}
F\left(\theta_s,\theta_i\right)=\exp\left(-\frac{\Delta k^2_x\sigma^2_x}{2}\right)\text{sinc}\left[\left(\Delta k_z+\Delta k_x\tan\rho\right)\frac{L}{2}\right].
\label{eq:TwoPhotonAmplitude}
\end{equation}
Here, the $z$ axis is assumed to be along the pump wavevector, $\Delta k_x = k_s\sin(\theta_s)+k_i\sin(\theta_i)$ is the transverse wavevector mismatch, $\sigma_x$ is the pump width (standard deviation of the gaussian field profile),  $\Delta k_z = k_p-k_s\cos(\theta_s)-k_i\cos(\theta_i)$ is the longitudinal wavevector mismatch, $L$ the length of the crystal, and $k_{p,s,i}$ are the pump, signal, and idler wavevectors. To obtain the wavelength-angular distribution of the signal beam intensity at low-gain PDC, the squared modulus of the TPA (\ref{eq:TwoPhotonAmplitude}) is integrated over all idler angles $\theta_i$ for each signal wavelength $\lambda_s$. The resulting intensity distribution, with the integration over idler and signal wavelengths, separately normalised, is shown in Fig. \ref{fig:SpectralAmplitudeLowGain}. One can see that even at low gain, the anisotropy manifests itself in the angular asymmetry of the spectrum: for instance, the upper branch is broader than the lower one, but its peak value is lower. This asymmetry has been described in a number of theoretical and experimental papers~\cite{DiLorenzoPires2011,Ramirez-Alarcon2013,Jeronimo-Moreno2014}.

In our experiment, the parametric gain is very high and hence the intensity distribution is modified.To describe this change of the TPA we start by employing the Schmidt decomposition of t he TPA~\cite{Sharapova:15},
\begin{equation}
F\left(\theta_s,\theta_i\right)=\sum_n\sqrt{\lambda_n}u_n(\theta_s)v_n(\theta_i).
\label{eq:SchmidtDecomposition}
\end{equation}
Here $u_n(\theta_s)$ and $v_n(\theta_i)$ are the Schmidt modes of the signal and idler radiation, respectively, and $\lambda_n$ are the Schmidt eigenvalues. At high parametric gain, the modes $u_n(\theta_s)$ and $v_n(\theta_i)$ are the same as at low gain, while the eigenvalues are redistributed and become~\cite{Sharapova:15}
\begin{equation}
\lambda'_n \propto \sinh^2\left(\Gamma\sqrt{\lambda_n}\right).
\label{eq:HighGain}
\end{equation}
The parameter $\Gamma$ can be found experimentally, see Fig. \ref{fig:Setup}, by measuring the output power $P_{PDC}$ of the generated PDC as a function of the input pump power $P$. The dependence has the form
\begin{equation}
P_{PDC}=A\sinh^2(B\sqrt{P}),
\label{eq:Gain}
\end{equation}
with A and B being fitting parameters, from which the single-mode parametric gain $G= B \sqrt{P}=\Gamma\sqrt{\lambda_0}$ is derived. To obtain the high-gain intensity distribution, it is sufficient to calculate the TPA (\ref{eq:SchmidtDecomposition}) with the new eigenvalues $\lambda'_n$. As a result, as is shown in Fig. \ref{fig:SpectralAmplitudeHighGain}, only two small regions of the distribution are amplified, one is the signal in the IR spectral range centred near the walk-off angle, and the other one is the idler in the visible range~\cite{Perez:13, Perez:14}.

\section{Experimental setup}
\begin{figure}[htbp]
\centerline{\includegraphics[width=\columnwidth]{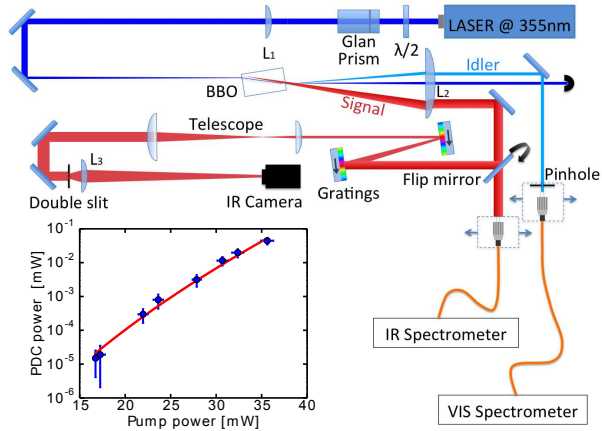}}
\caption{Experimental setup. The lens L$_1$ (focal length 50, 75, or 100 cm and the position changed accordingly) focuses the beam on the crystal. Another lens L$_2$, with 10 cm focal length and a 2'' diameter, collects all the generated radiation. The lens L$_3$ has a focal length of 50 cm and an IR camera is placed in its focal plane. Inset: PDC output power versus the input pump power, measured with 80 \textmu m pump waist. Solid line: the fit with Eq.~(\ref{eq:Gain}).}
\label{fig:Setup}
\end{figure}
The experimental setup is shown in Fig. \ref{fig:Setup}. A 10 mm BBO crystal was pumped by the third-harmonic radiation of a Nd:YAG laser at 355 nm wavelength, 1 kHz repetition rate and 18 ps pulse width. A half-wave plate and a Glan-Thomson prism polariser were used to ensure the correct polarisation of the beam. By using different focusing lenses, it was possible to change the pump beam waist inside the crystal. We used three different lenses with focal lengths of 500, 750 and 1000 mm which corresponded to  80, 130 and 160 \textmu m FWHM waists, respectively. The PDC radiation was then generated under the non-collinear phase matching configuration. The signal in the near infrared (NIR) range was generated at the walk-off angle while the idler was generated in the conjugated direction. Both beams were collected by a collimating lens with a focal length of 10 cm while the remaining pump radiation was blocked. The spectra were measured by scanning the beams with fibres coupled to spectrometers with 50 \textmu m spatial resolution. The fibre tips were positioned in the focal plane of the collimating lens to provide far-field intensity distributions.

In the infrared arm, a flip mirror was placed that reflected the beam towards the beam shaping setup. Here the quality of the beam was improved by means of two blazed diffraction gratings, which compensated for the wavelength-angle dependence (angular chirp) of PDC and overlapped all wavelength components~\cite{Katamadze:15}. Both gratings were reflective with 600 lines per mm. The first one converged all wavelength components while the second one, placed at the converging point, ensured parallel propagation of all wavelengths. The gratings were followed by a cylindrical lens telescope reducing the beam ellipticity. For instance, in the case of the pump waist 130 \textmu m the beam had to be magnified three times in the horizontal direction, which was achieved by using cylindrical lenses with focal lengths of 10 cm and 30 cm.

After the beam shaping, the measurement of the spatial coherence was performed. This was done by placing a double slit in the beam and, with the help of a lens, measuring the interference fringes in the far field with an IR CCD camera. The measurement was repeated with many set of slits each of 0.4 mm width and with different spacings, ranging between 0.2 to 6 mm.

\section{Results}
In order to evaluate the parametric gain, the output PDC power was measured versus the input pump power and then fitted by Eq.~(\ref{eq:Gain}).
The measurement was performed with the IR beam, for all three available pump waists. As expected, for smaller waists the gain is higher: at 35 mW pumping, $G=10.5\pm0.5$ for 80 \textmu m waist (Fig. \ref{fig:Setup}), $G=7.0\pm0.8$ for 130 \textmu m and $G=5.9\pm0.6$ for 160 \textmu m. The fit is valid only in the non-depleted pump regime, which was the case only for low pump power. The maximum conversion efficiency observed was 24\% and was obtained with a waist of 130 \textmu m and a pump power of 51 mW. Under these conditions, the output pump was strongly depleted.

\begin{figure}[h]
\begin{subfigure}[h]{0.49\columnwidth}
\includegraphics[width=\textwidth]{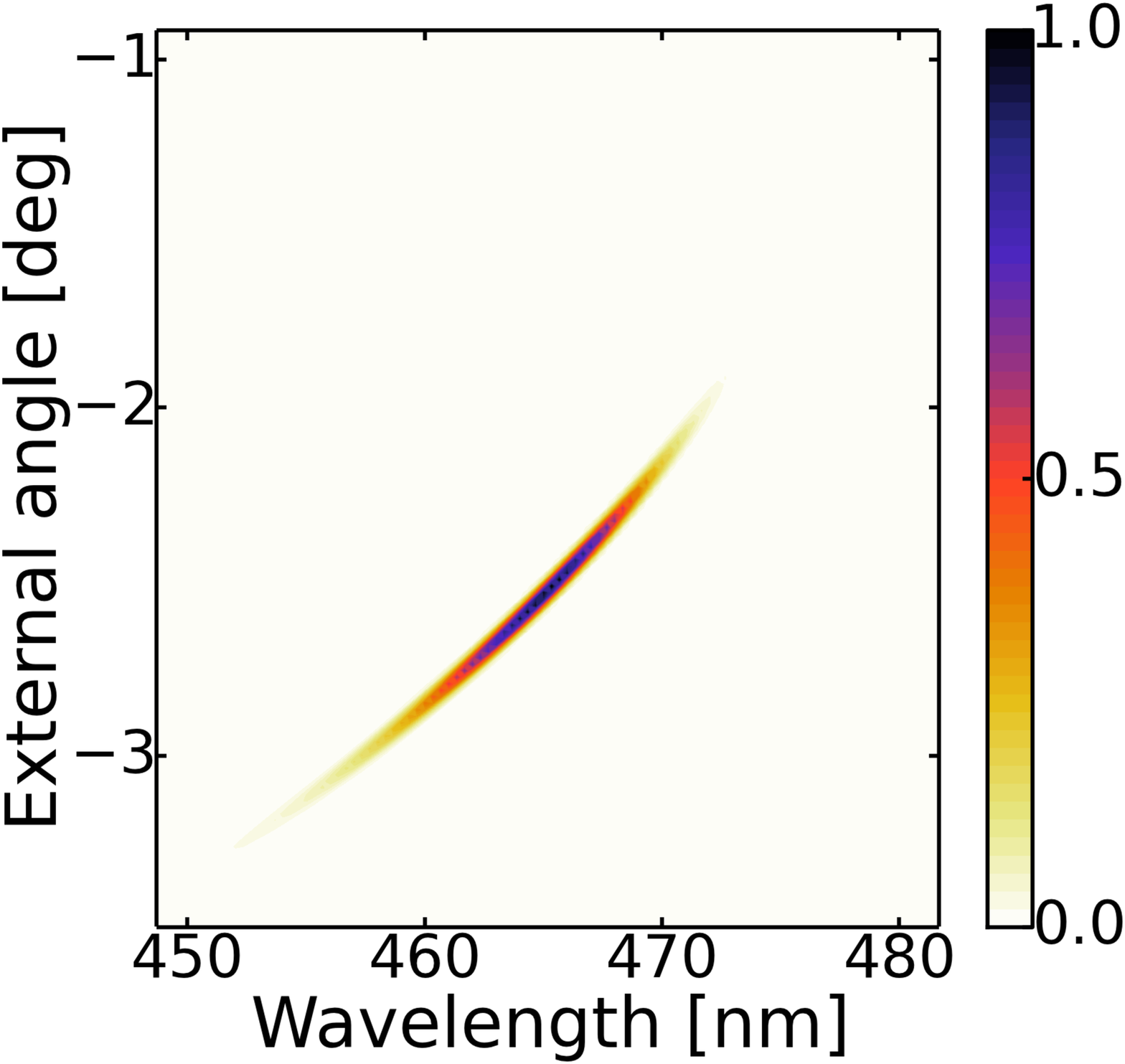}
\caption{}
\label{fig:SpectralAmplitudeC750VIS}
\end{subfigure}
\begin{subfigure}[h]{0.49\columnwidth}
\includegraphics[width=\textwidth]{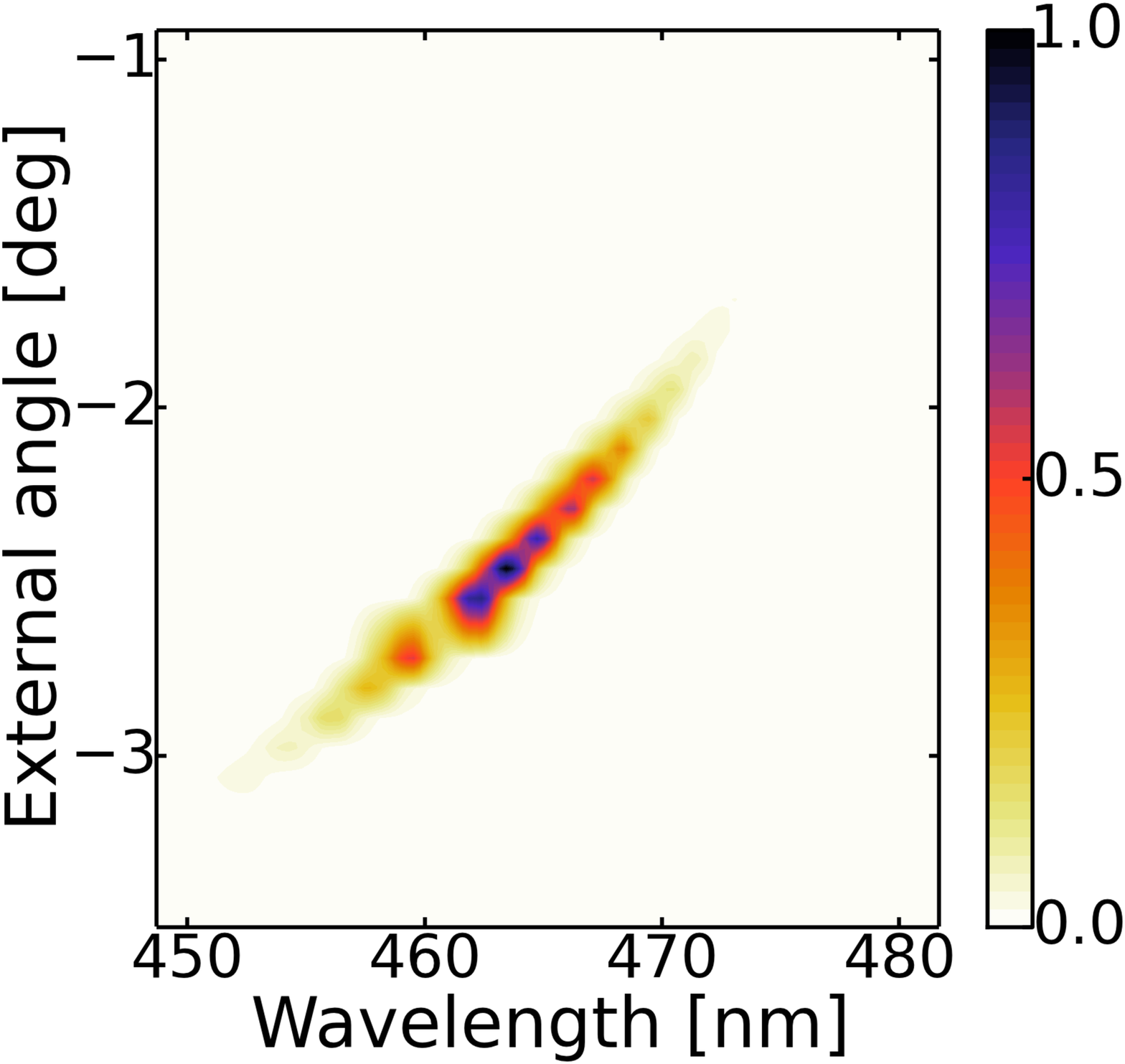}
\caption{}
\label{fig:SpectralAmplitudeM750VIS}
\end{subfigure}
\begin{subfigure}[h]{0.49\columnwidth}
\includegraphics[width=\textwidth]{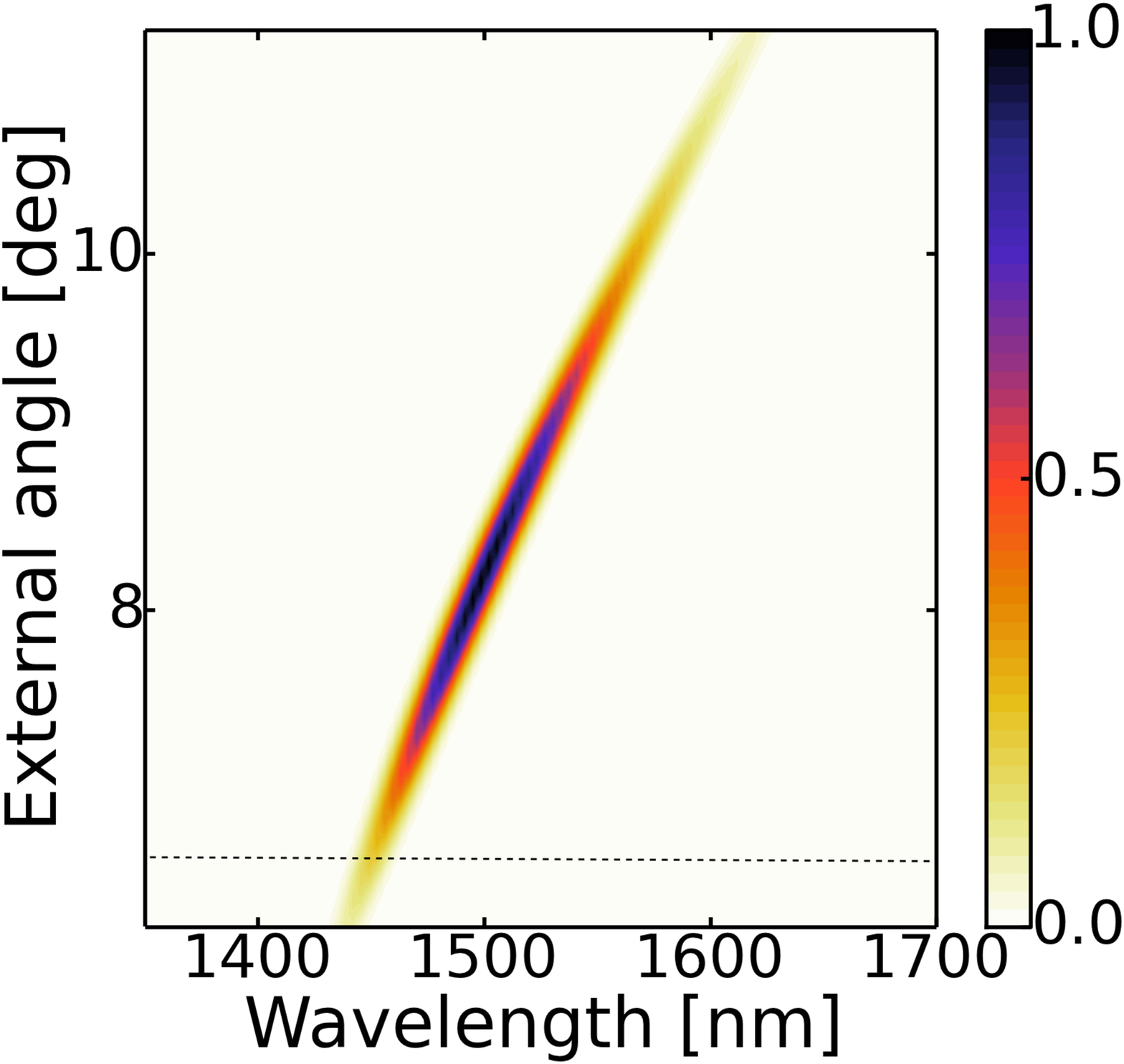}
\caption{}
\label{fig:SpectralAmplitudeC750}
\end{subfigure}
\begin{subfigure}[h]{0.49\columnwidth}
\includegraphics[width=\textwidth]{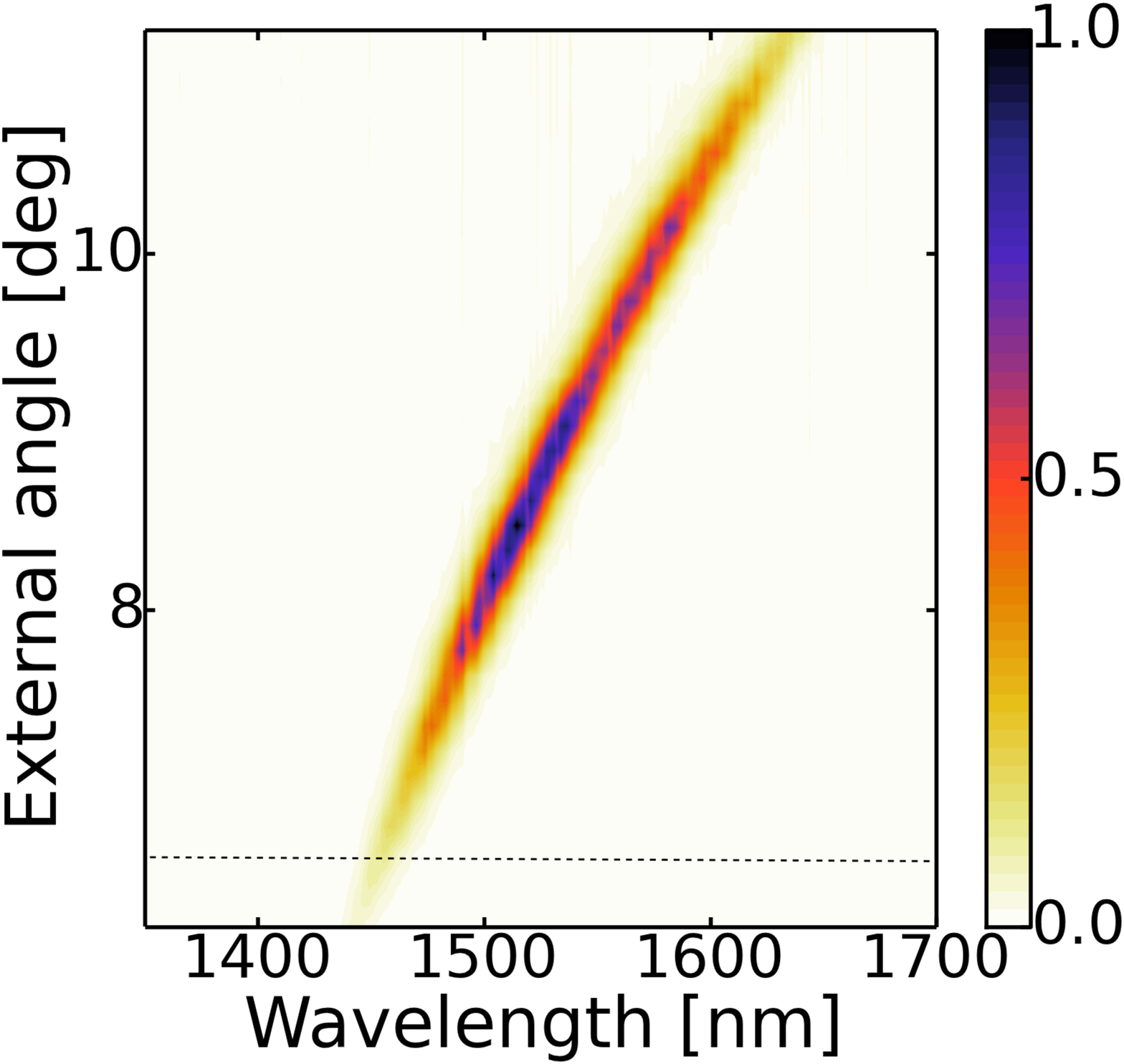}
\caption{}
\label{fig:SpectralAmplitudeM750}
\end{subfigure}
\caption{Wavelength-angular spectrum, calculated (a,c) and measured (b,d), of the idler (a,b) and signal (c,d) beams generated with the 130 \textmu m pump waist.}
\end{figure}

Next, we analyse the wavelength-angular spectrum of the source. The measurements were performed for all three available pump waists. The wavelength-angular spectra of the signal and idler beams are shown in Figs. \ref{fig:SpectralAmplitudeM750VIS}, \ref{fig:SpectralAmplitudeM750}, together with the corresponding numerically calculated spectra (\ref{fig:SpectralAmplitudeC750VIS},~\ref{fig:SpectralAmplitudeC750}).
In the calculation, the length of the crystal is taken $L=5$ mm in order to take into account the effect of the temporal walk-off due to the different group velocities of the signal and the pump. For a 10 mm crystal the group delay between the beams is 3.7 ps that is on the order of the coherence time of the laser. 

One can notice that the maximum of the emission corresponds not exactly to the walk-off angle (shown by dashed line) but to a slightly larger angle.  This is probably due to a narrower angular bandwidth and, accordingly, a higher peak intensity, at larger wavelengths, leading to the shift of the maximum towards larger angles.

\begin{figure}[htbp]
\begin{subfigure}[h]{0.49\columnwidth}
\includegraphics[width=\textwidth]{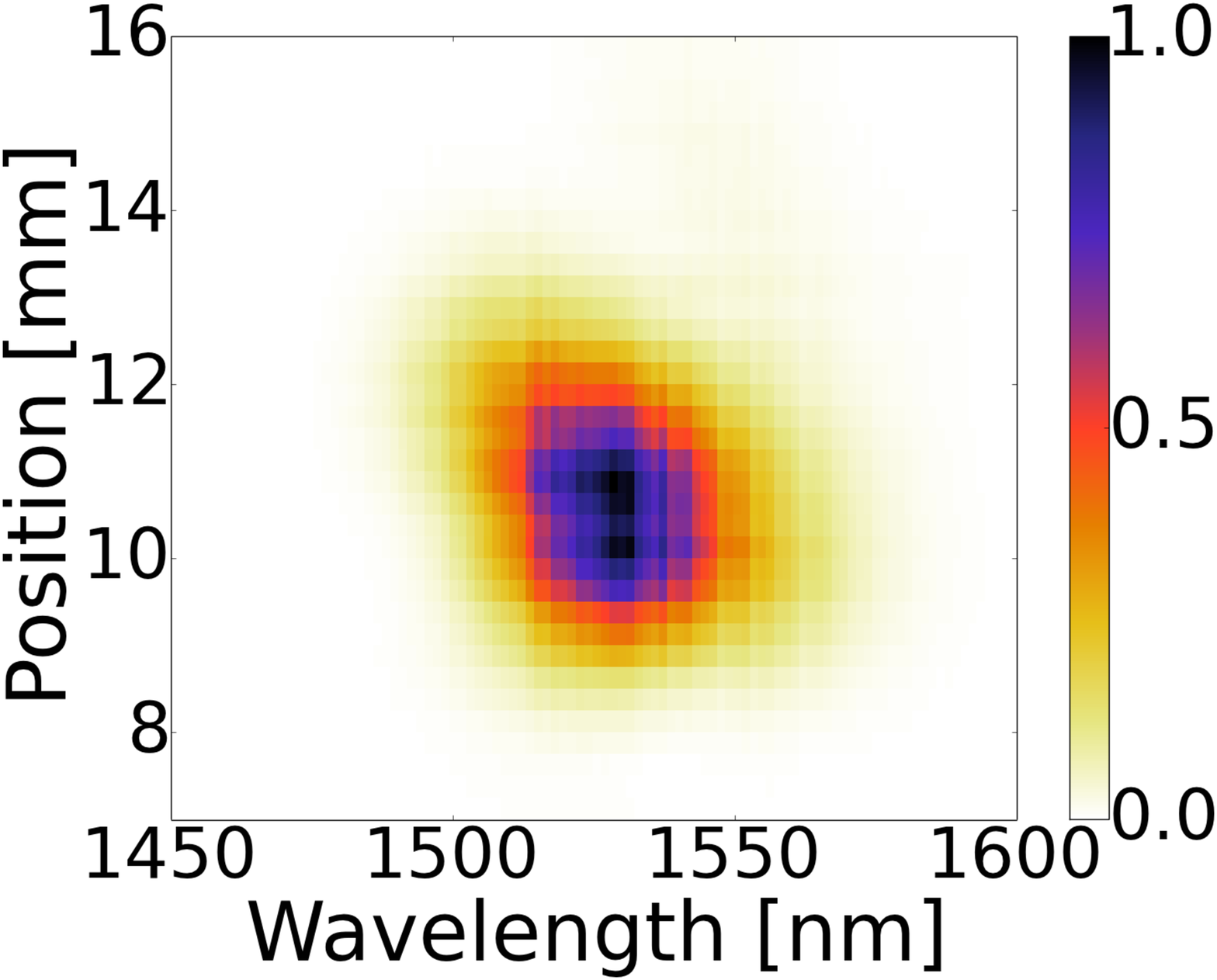}
\caption{}
\label{fig:SpectralAmplitudeAfterGratings}
\end{subfigure}
\begin{subfigure}[h]{0.49\columnwidth}
\includegraphics[width=\textwidth]{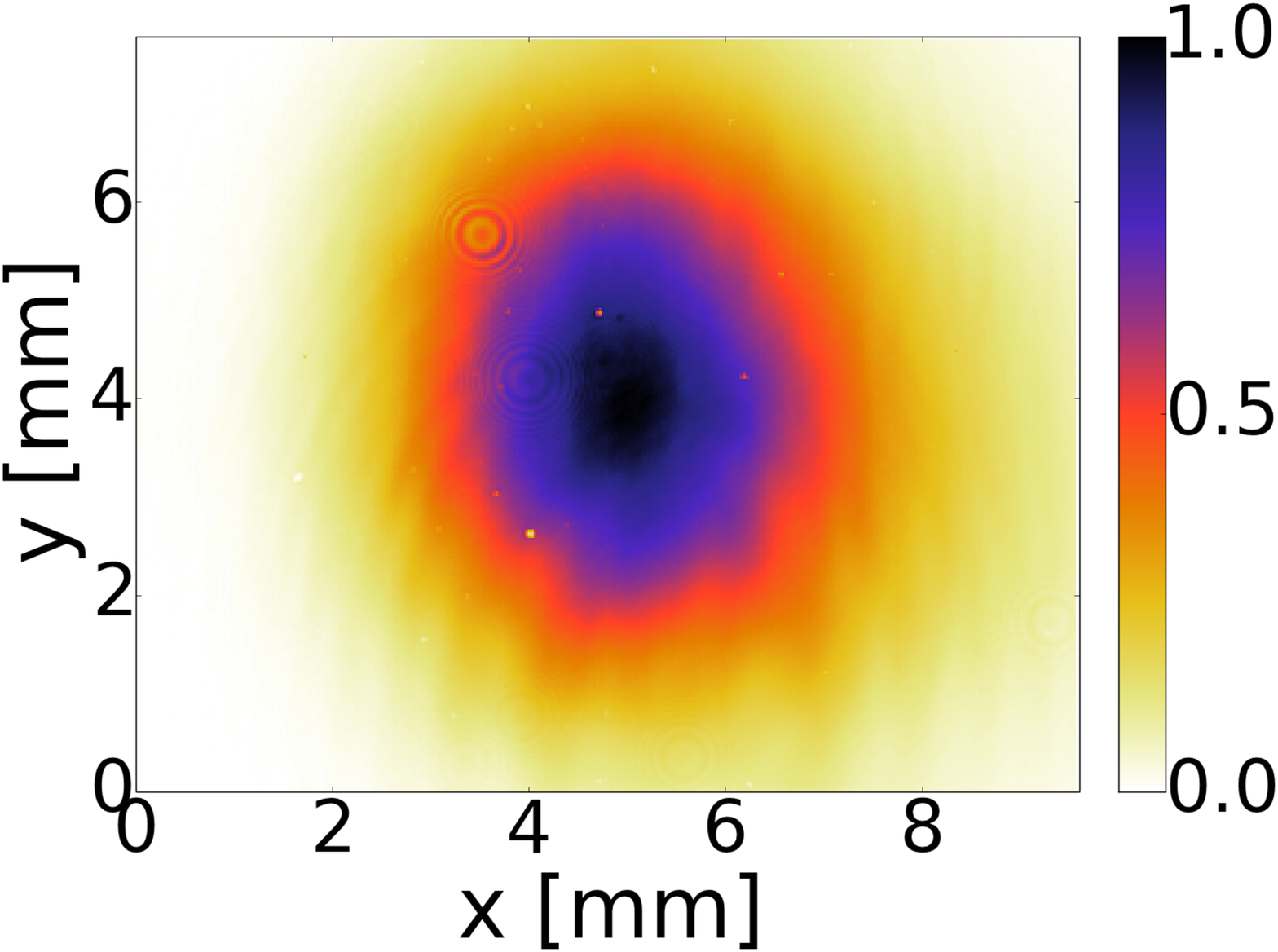}
\caption{}
\label{fig:BeamShape50}
\end{subfigure}
\caption{Wavelength-angular spectrum of the signal after the beam shaping (a) and the spatial intensity distribution of the optimised beam generated with 130 \textmu m pump waist (b).}
\end{figure}

Using the gratings, it is then possible to eliminate the angular chirp and to combine all wavelengths into a single beam  (Fig.~\ref{fig:SpectralAmplitudeAfterGratings}).

The final step is the measurement of the spatial coherence of the beam and, in particular, the number of spatial modes. For many applications a single mode is preferable, for instance whenever a single mode fibre has to be used.
In order to estimate the number of modes we study the spatial coherence function $G^{(1)}(x,x')$ \cite{Glauber:63}. In a Young's double-slit experiment, the visibility of interference for the slits placed at positions -x and x is given by $G^{(1)}(-x,x)$. As the spacing between the slits is increased, the visibility reduces, the coherence radius corresponding to the spacing for which the fringes visibility drops to 50\% \cite{Klyshko:11}. Using different pairs of slits, it is possible to map the anti-diagonal distribution of $G^{(1)}(-x,x)$. On the other hand, the beam profile corresponds to the diagonal distribution, $G^{(1)}(x,x)$. Assuming the Gaussian Schell's model \cite{Schell:61}, it is possible to fully reconstruct the coherence function:
\begin{equation}
G^{(1)}\left(x,x'\right)=\exp\left[-\frac{(x+x')}{2a^2}\right]\exp\left[-\frac{(x-x')}{2b^2}\right],
\label{eq:correlation}
\end{equation}
where the standard deviations $a$ and $b$ are calculated from the beam profile and the visibility distributions, respectively. It is possible to apply Mercer's decomposition to this function~\cite{Mandel:95},
\begin{equation}
G^{(1)}(x,x')=\sum_ns_n\phi_n(x)\cdot\phi^*_n(x')
\label{eq:Mercer}
\end{equation}
and, from the eigenvalues $s_n$, with $\sum_n s_n=1$, one can calculate the number of modes $M_x=\sum_n1/ s_n^2$.

\begin{figure}[htb]
\includegraphics[width=\columnwidth]{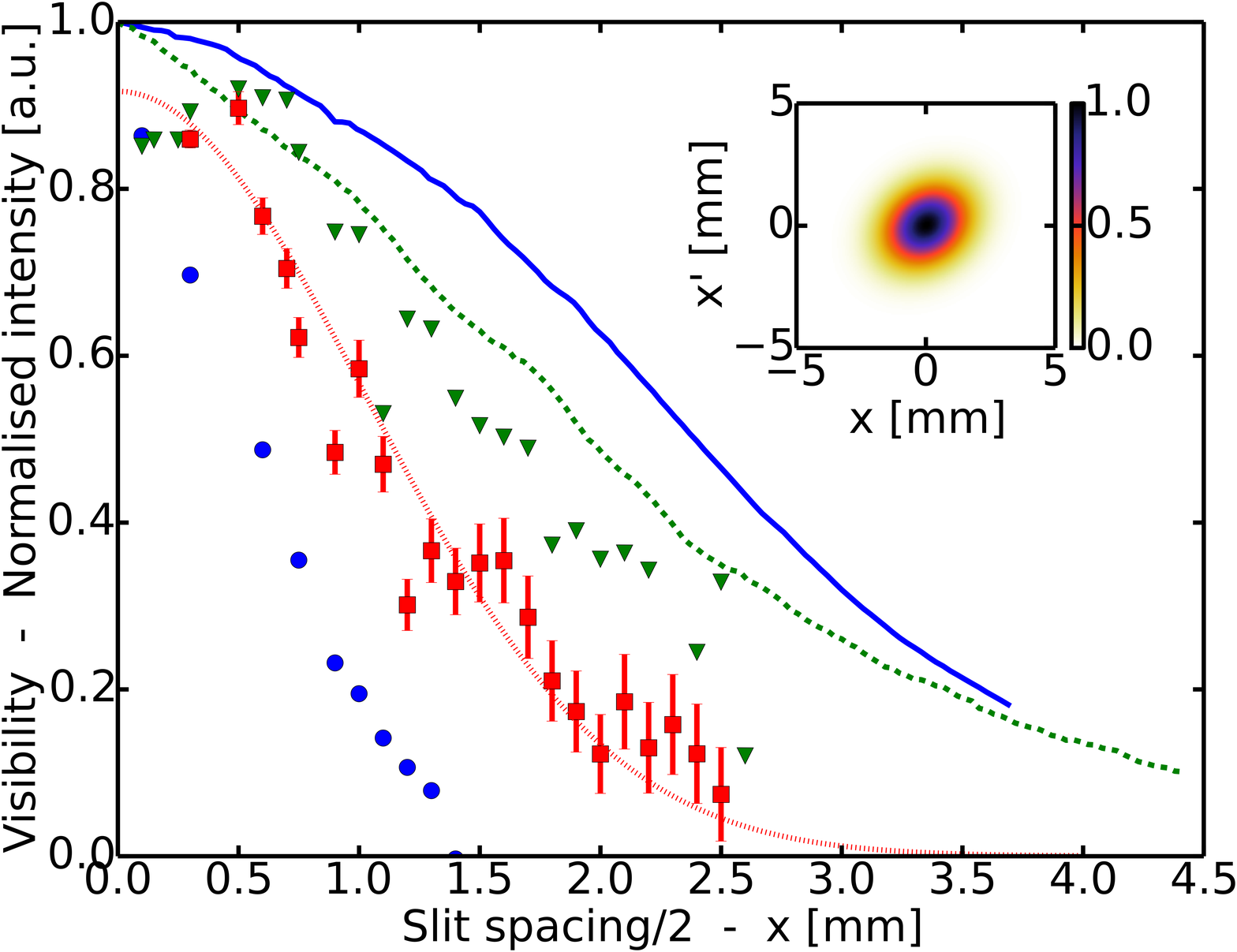}
\caption{The measured horizontal beam profile (green dashed line), the vertical beam profile (blue continuous line) and the visibility for horizontal displacement of the slits without bandpass filter (green triangles) and for vertical displacement of the slits with (red squares) and without (blue circles) bandpass filter of 12 nm for 130 \textmu m pump waist. The red dotted line represents the gaussian fit of the square data points. The inset shows the $G^{(1)}$ function for the horizontal direction calculated using the Schell model.}
\label{fig:Interference}
\end{figure}

The number of modes $M_x$ corresponds only to one direction as the subscript indicates. The total number of spatial modes in the beam is given by the product of the mode numbers in two orthogonal directions,
\begin{equation}
M_{tot}=M_xM_y.
\label{eq:NumberModes}
\end{equation}
 Fig. \ref{fig:Interference} shows the results obtained with the 130 \textmu m pump waist. Without any frequency filtering the total number of modes is measured to be $M_{tot}=2.02$ while including a bandpass filter of 12 nm for the vertical displacement measurement reduces the number of modes to 1.32. Although this was not tested in this work, a single mode of signal or idler PDC radiation is known to have thermal statistics~\cite{Boitier2011,Iskhakov2012}, which makes it very efficient for multi-photon effects like optical harmonic generation or multiphoton absorption~\cite{Jechow2013}. Therefore, the proposed source will be useful for nonlinear optics provided that the intensity fluctuations are not suppressed by pump depletion.

\section{Conclusion}
We have analysed the spectral and coherence properties of an OPG based on high-gain PDC generated along the pump Poynting vector. The source is very efficient and, in our configuration, provides up to 24\% of energy conversion. The radiation is spatially coherent and frequency-tuneable within the whole transparency range of the crystal. However, in this paper the analysis is concentrated around telecom wavelengths that are of great interest for optical fibre technologies.
Finally, due to the nature of the PDC process, the signal beam always emerges together with the idler beam, which is its copy in terms of the photon number and is anti-correlated in frequency. Thanks to these correlations it is possible, for example, to infer the signal beam properties by measuring the idler beam. This is especially important if the former is in a spectral range not accessible with measuring devices. The research leading to these results has received funding from the EU FP7 under grant agreement No. 308803 (project BRISQ2). We are grateful to Olga Tikhonova for helpful discussions.

\bibliographystyle{ieeetr}
\bibliography{PaperGPDCRef}

\end{document}